\documentclass[conference]{IEEEtran}

\IEEEoverridecommandlockouts

\usepackage{amsmath,amsthm,amsfonts}
\usepackage{cite}
\usepackage{bm}
\usepackage{tikz}
\usepackage{pgfplots,pgfplotstable}
\usetikzlibrary{calc,arrows,arrows.meta}
\usepackage{siunitx}

\usepackage{algorithmic}

\ifCLASSOPTIONcompsoc
  \usepackage[caption=false,font=normalsize,labelfont=sf,textfont=sf]{subfig}
\else
  \usepackage[caption=false,font=footnotesize]{subfig}
\fi

\usepackage{url}

\newcommand{\av}{\bm{a}}
\newcommand{\hv}{\bm{h}}
\newcommand{\pv}{\bm{p}}
\newcommand{\sv}{\bm{s}}
\newcommand{\uv}{\bm{u}}
\newcommand{\wv}{\bm{w}}
\newcommand{\xv}{\bm{x}}
\newcommand{\yv}{\bm{y}}
\newcommand{\zv}{\bm{z}}
\newcommand{\Am}{\bm{A}}
\newcommand{\Rm}{\bm{R}}
\newcommand{\Wm}{\bm{W}}

\newcommand{\Exp}{{\mathbb{E}}}   
\newcommand{\Tr}{{\mathsf{T}}}   

\begin{document}
\title{Neural Network Aided Decoding for\\Physical-Layer Network Coding Random Access}

\author{\IEEEauthorblockN{Adriano Pastore\IEEEauthorrefmark{1},
Paul de Kerret\IEEEauthorrefmark{2},
Monica Navarro\IEEEauthorrefmark{1}, 
David Gregoratti\IEEEauthorrefmark{1},
and David Gesbert\IEEEauthorrefmark{2}}
\vspace{2ex}
\IEEEauthorblockA{\IEEEauthorrefmark{1}CTTC/CERCA, Avinguda Carl Friedrich Gauss 7, 08860 Castelldefels, Spain} 
\IEEEauthorblockA{\IEEEauthorrefmark{2}EURECOM, 450 route des Chappes, 06410 Biot, Sophia Antipolis, France} 
\thanks{This work was supported by the Catalan Government under grant 2017 SGR 1479 and by the Spanish Ministery of Economy and Competitiveness through projects TEC2014-59255-C3-1-R (ELISA) and TEC2015-69868-C2-2-R (ADVENTURE). D.\ Gesbert and P.\ de Kerret are supported by the European Research Council under the European Union's Horizon 2020 research and innovation program (Agreement no.\ 670896).}
}

\IEEEspecialpapernotice{(Invited Paper)}
\maketitle

\begin{abstract}
Hinging on ideas from physical-layer network coding, some promising proposals of coded random access systems seek to improve system performance (while preserving low complexity) by means of packet repetitions and decoding of {\em linear combinations} of colliding packets, whenever the decoding of individual packets fails. The resulting linear combinations are then temporarily stored in the hope of gathering enough linearly independent combinations so as to eventually recover all individual packets through the resolution of a linear system at the end of the contention frame. However, it is unclear which among the numerous linear combinations---whose number grows exponentially with the degree of collision---will have low probability of decoding error. Since no analytical framework exists to determine which combinations are easiest to decode, this makes the case for a machine learning algorithm to assist the receiver in deciding which linear combinations to target. For this purpose, we train neural networks that approximate the error probability for every possible linear combination based on the estimated channel gains and demonstrate the effectiveness of our approach by numerical simulations.
\end{abstract}

\IEEEpeerreviewmaketitle

\section{Introduction}

With the increasing importance of machine-to-machine communication in the context of the Internet of Things (IoT), many {\em ad hoc} wireless networks are witnessing a drastic densification. In typical massive machine-to-machine communication settings, numerous terminals with bursty traffic demands and scarce coordination seek to convey information packets to a central node. Early concepts of random access such as the legacy ALOHA scheme~\cite{Ab70,Ro75} were, like most medium access schemes in packet-switched networks, based on the paradigm of collision avoidance~\cite{KlSc80,TaJaBa04}. However, the more crowded the wireless medium, the more necessary it becomes to devise protocols that are capable of resolving collisions rather than discarding them. In addition, harsher delay constraints and limitations on feedback capabilities in many IoT-related applications (Industry 4.0, vehicle-to-vehicle communication, etc.) make retransmissions impractical or costly at best. As a result, rather than treating collisions as losses, modern proposals of grant-free random access protocols have evolved towards the use of more sophisticated forward error correction techniques, often referred to as multipacket reception (MPR)~\cite{GhVeSc88,ToZhMe01,NaMeTo05,LuEp06,ChBeTo13,BaChAl14}.

Of notable importance among well-known MPR techniques is {\em successive interference cancellation} (SIC) for its low complexity as compared to joint decoding: when one user experiences a substantially stronger channel gain than others, its packets can be decoded and subtracted (cancelled) from the received signal~\cite{Li11,ZaZo12}, 
thereby potentially allowing more subsequent decoding operations or conflict resolutions. Liva's work~\cite{Li11} on Contention Resolution Diversity Slotted ALOHA (CRDSA) shows that the contention resolution with SIC is analogous to iterative belief-propagation (BP) erasure decoding and can thus be built upon established BP code designs. In the context of ultrareliable low-latency communications, \cite{StLaPo17} characterizes the reliability--latency performance of frameless ALOHA with SIC decoding. 

Beyond SIC, another promising strategy, which is inspired by physical-layer network coding (PLNC), consists in encoding and decoding {\em linear combinations} of packets~\cite{CoPf14,CoPfNa16,GoGaWe15,AsFeRoKs15}. By collecting enough linearly independent packet combinations within a contention period (frame), the receiver might be able to resolve all individual packets. The motivation behind this approach is that, where SIC fails, the decoding of linear combinations might still succeed and unlock useful information for contention resolution. As another advantage over CRDSA, this PLNC random access scheme effectively implements some form of fast frame-level contention resolution, yet without the need of storing raw receive signals until the end of the contention period. The authors of~\cite{GoStPo14,AsFeRo17} also verified a notable performance advantage of PLNC random access over purely SIC based schemes.

However, given the large number of possible linear combinations---which is exponential in the number of colliding packets---and the fact it is unclear {\em a priori} which ones can be reliably decoded, the brute-force approach of attempting to decode them all is prohibitive. In this article, we propose to enhance the decoding architecture with a machine learning algorithm based on deep neural networks (DNN) that determines the linear combinations that can be most reliably decoded. Such an algorithm is a necessary step in making PLNC random access practically feasible.

\section{System description}

\subsection{PLNC coded random access scheme}

We focus on the coded random access scheme from~\cite{CoPfNa16}, which was adapted and proposed for massive access connectivity within European project~\cite{D41,D42}. We generally adopt the corresponding numerology, though all considerations in this paper can be straightforwardly extended to different parameter values.

Consider an $L$-user slotted random access channel with a user index set $[L] = \{1,\dotsc,L\}$. All signals are in real-valued baseband representation and indexed by discrete time indices $t \in \mathbb{Z}$ which are structured as follows: a {\em slot} is a group of $T_\mathsf{s} = 168$ consecutive time instants, whereas a {\em frame} is a group of $T_\mathsf{f} = 10$~consecutive slots. We assume that a control loop allows the terminals to time their transmissions in such way that the received signals are precisely synchronized on symbol, slot and frame level.

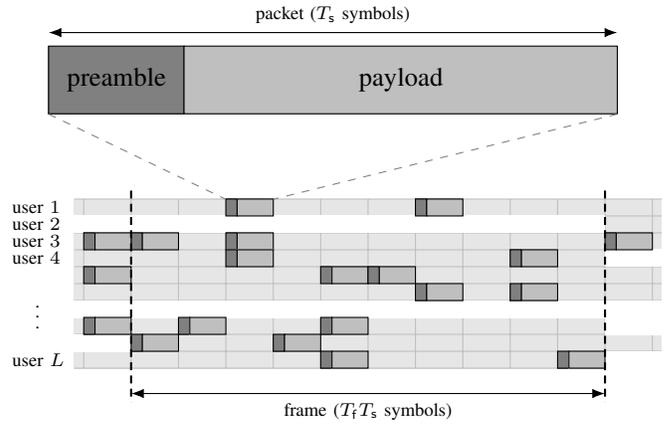
\begin{figure}[h]
\begin{center}

\newcommand\packet[2]{
\begin{scope}[shift={(#1,-#2)}]
	\draw[fill=black!50] (0,0) rectangle (.2381,1);
	\draw[fill=black!25] (.2381,0) rectangle (1,1);
\end{scope}
}

\begin{tikzpicture}[remember picture, scale=.9]
	\begin{scope}[x=8.4cm, y=1cm]
		\draw[fill=black!50] (0,0) rectangle node {preamble} (.2381,1);
		\draw[fill=black!25] (.2381,0) rectangle node {payload} (1,1);
		\coordinate (big left) at (0,0);   
		\coordinate (big right) at (1,0);   
		\draw[Latex-Latex] (0,1.2) -- node[above] {\scriptsize packet ($T_\mathsf{s}$ symbols)} (1,1.2);
	\end{scope}
\end{tikzpicture}

\vspace{1cm}

\begin{tikzpicture}[remember picture, scale=.75, x=.84cm, y=.3cm]
	\coordinate (small left) at (2,0);
	\coordinate (small right) at (3,0);
	\fill[black!10] (-1.2,0) rectangle (11.2,-10);
	\draw[black!25, step=1] (-1.2,0) grid (11.2,-10);
	\fill[white] (0,-6) rectangle +(10,-1);   
	\fill[white] (0,-1) rectangle +(10,-1);   
	\fill[white] (-1.2,-1) rectangle (0,-2);   
	\fill[white] (-1.2,-6) rectangle (0,-7);   
	\fill[white] (-1.2,-8) rectangle (0,-9);   
	\fill[white] (10,-6) rectangle (11.2,-7);   
	\fill[white] (10,-7) rectangle (11.2,-8);   
	\fill[white] (10,-9) rectangle (11.2,-10);   
	\fill[white] (10,-3) rectangle (11.2,-4);   
	\packet{2}{1} \packet{6}{1}
	\packet{0}{3} \packet{2}{3}
	\packet{2}{4} \packet{8}{4}
	\packet{4}{5} \packet{5}{5}
	\packet{6}{6} \packet{8}{6}
	\packet{1}{8} \packet{4}{8}
	\packet{0}{9} \packet{3}{9}
	\packet{4}{10} \packet{9}{10}
	\packet{-1}{3}
	\packet{-1}{5}
	\packet{-1}{8}
	\packet{10}{3}
	\draw[thick, densely dashed] (0,.5)--(0,-12) (10,.5)--(10,-12);
	\draw[Latex-Latex] (0,-11.5) -- node[below] {\scriptsize frame ($T_\mathsf{f} T_\mathsf{s}$ symbols)} (10,-11.5);
	\foreach \i/\y in {1/-0.5,2/-1.5,3/-2.5,4/-3.5,L/-9.5}{
		\node[name=user\i, anchor=west] at (-2.7,\y) {\scriptsize user~$\i$};
	}
	\node at ($(user4)!.5!(userL)$) {\scriptsize \vdots};
\end{tikzpicture}

\begin{tikzpicture}[remember picture, overlay]
	\draw[thin, dashed, gray] (big left)--(small left) (big right)--(small right);
\end{tikzpicture}

\end{center}
\caption{Exemplary illustration of the slot and frame traffic pattern in the PLNC coded random access system under study. For this picture, we have set $T_\mathsf{f} = 10$ and $r = 2$. Every row in the above grid-like structure depicts the activity of one user over time. Frames in which a user is idle are blanked out.}
\label{fig:transmission_pattern}
\end{figure}

In each frame, a random subset $\mathcal{A} \subseteq [L]$ of so-called {\em active} users attempt to convey their payload data, while other users remain silent. Each active user is set to transmit $r$ replicas of a {\em packet}\footnote{The number $r$ of packet repetitions can be optimized in line with~\cite{Li11}.} in $r$~randomly chosen slots out of the $T_\mathsf{f}$ slots that compose the frame, while remaining silent in other slots.\footnote{We talk of {\em coded} random access due to this repetition coding, which the receiver can exploit at frame level to resolve collisions.} A packet consists of a preamble of length $T_\mathsf{p}=40$ symbols that carry a unique user signature that serves simultaneously for user identification and channel estimation\footnote{In general, the tasks related to the preamble processing (multiuser detection, user identification, collision detection, channel estimation) impact overall performance and resource allocation strategies. Thus, some works treat them as an essential part of the random access scheme
. For simplicity however, in this work we assume that they are {\em genie-aided}, so we can focus entirely on payload decoding and leave a joint detection--decoding perspective for future work.} (e.g., via orthogonal matching pursuit at the receiver), and a payload of length $n=128$, which contains the codeword to be transmitted. As a result, in each slot, a random subset $\mathcal{T} \subseteq \mathcal{A}$ of active users will be simultaneously transmitting their packets.\footnote{Hence $\mathcal{T}$ varies from slot to slot, and $\mathcal{A}$ from frame to frame, yet we refrain from specifying the slot/frame indices so as to not encumber notation.} We say that $|\mathcal{T}|$ is the {\em collision degree} in that slot, where $\left|\cdot\right|$ denotes set cardinality. The resulting transmission pattern is depicted in Figure~\ref{fig:transmission_pattern}.

At time instant $t \in \mathbb{Z}$ pertaining to a slot in which users $\mathcal{T}$ are transmitting payload data, we have that, conditioned on transmit symbols $X_\ell = x_\ell \in \mathbb{R}$, $\ell \in \mathcal{T}$, the channel output is given by
\begin{equation}
	Y(t)
	= \sum_{\ell \in \mathcal{T}} h_\ell x_\ell(t) + Z(t)
\end{equation}
where $Z(t) \sim \mathcal{N}(0,1)$ is additive white Gaussian noise. We assume BPSK signaling with power $P$, hence $X_\ell \in \{-\sqrt{P},+\sqrt{P}\}$.
The channel gains $h_\ell \in \mathbb{R}$ are drawn from some random distribution (to be specified) and are assumed to stay constant over the duration of a slot, but they vary independently from one slot to the next.

\subsection{Coding and modulation}

All users employ the same rate-$1/2$ short LDPC code with message size $k=64$ bits and codeword length $n=128$ bits~\cite{CCSDS}. Since linear binary codebooks are closed under modulo-$2$ addition, any linear combination
\begin{equation}
	\wv_{\av}
	= \bigoplus_{\ell \in \mathcal{T}} a_\ell \uv_\ell
\end{equation}
of codewords $\uv_\ell \in \{0,1\}^n$ (as row vectors) with binary coefficients $a_\ell \in \{0,1\}$, $\ell \in \mathcal{T}$, belongs to the codebook. The received payload signal can be expressed in vector notation as
\begin{equation}
	\yv
	= \sum_{\ell \in \mathcal{T}} h_\ell \xv(\uv_\ell) + \zv.
\end{equation}
The modulation mapping $\xv(\uv_\ell) = \bigl[ x(u_{\ell,1}), \dotsc, x(u_{\ell,n}) \bigr]$ is such that code bits are mapped to BPSK symbols as $x(0)=-\sqrt{P}$ and $x(1)=+\sqrt{P}$.

\subsection{Decoding}

The decoding procedure is performed at slot level and at frame level.

\subsubsection{Slot-level decoding}

The slot-level decoding is performed in two successive stages:

\paragraph{Successive interference cancellation (SIC)}

Within each time slot, based on a vector of channel gains $\hv$ and a receive vector $\yv$, the slot-level decoder attempts to recover as many individual codewords $\uv_\ell$ as possible by means of BP decoding. The users are decoded in order of decreasing channel gains and their interference is successively subtracted from the observation $\yv$.

\paragraph{Sum decoding (SD)}

\begin{figure}[ht]
\begin{center}
\begin{tikzpicture}
\pgfplotstableread{
label	iteration1	iteration2	iteration3
1	152303	478	2
2	107905	6803	19
3	88421	2955	4
4	54579	1028	0
5	17198	140	0
6	2242	0	0
}\countdata
\begin{axis}[
	y = 5mm,
	xbar stacked,
	bar width = 3mm,
	xmin = 0,
	enlarge y limits = .1,
	legend style = {at = {(.95,.95)}, anchor=north east, legend columns=1},
	xlabel = {counts of successful decoding},
	ylabel = {degree of\\linear combination},
	ylabel style = {align=center},
	ytick = data,
	y tick label style={anchor=east},
	axis on top,
]
\addplot[xbar, draw=none, fill=black!30] table [y=label, x=iteration1] {\countdata};
\addplot[xbar, draw=none, fill=black!60] table [y=label, x=iteration2] {\countdata};
\addplot[xbar, draw=none, fill=black] table [y=label, x=iteration3] {\countdata};
\legend{iteration 1, iteration 2, iteration 3}
\end{axis}
\end{tikzpicture}
\end{center}
\caption{Counts of successful decoding for slot-level decoding of $|\mathcal{T}|=6$ colliding packets, based on a simulation with \num{400000}~slots, user-symmetric Rayleigh fading and average SNR of~\SI{15}{\decibel}. The {\em degree of linear combination} (ordinate axis label) stands for the Hamming weight of $\av$.}
\label{fig:counts}
\end{figure}
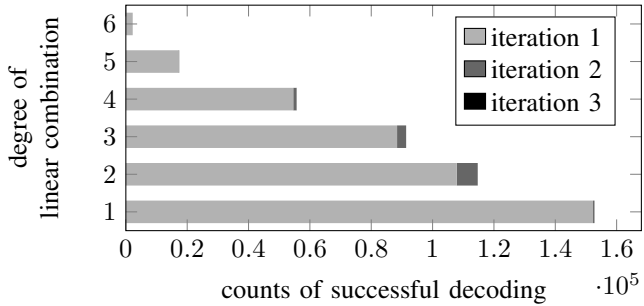

Suppose that after some SIC iterations, a subset $\mathcal{T}' \subseteq \mathcal{T}$ of packets remains undecoded and the BP decoding fails for the next packet in line. Based on the channel coefficients $\hv$ and the remaining signal $\yv' = \yv - \sum_{\ell \in \mathcal{T} \setminus \mathcal{T}'} h_\ell \xv(\hat{\uv}_\ell)$, the sum decoder now attempts to recover as many codeword combinations $\wv_{\av}$ as possible, where the weight vector $\av = \bigl[ a_1,\dots,a_{\mathcal{T}'} \bigr] \in \{0,1\}^{|\mathcal{T}'|}$ is varied through all possibilities. There are $2^{|\mathcal{T}'|} - |\mathcal{T}'| - 1$ such weight vectors to be considered, since we exclude the all-zero vector and the $|\mathcal{T}'|$ singleton vectors (which are not decodable due to the SIC routine having terminated). This sum decoding is done by a variation of the BP decoder (see Section~4.4 in~\cite{CoPfNa16} and Section~III in~\cite{Pf14} for details).
The usefulness of sum decoding is best visualized by the histogram in Figure~\ref{fig:counts}, which makes it evident that even when SIC fails, a large amount of codeword combinations can still be reliably decoded and are potentially helpful for frame-level decoding.

In the experiment from Figure~\ref{fig:counts}, the receiver proceeds as follows: in the first iteration, it exhaustively attempts to decode all $2^{|\mathcal{T}|}-1=63$ combinations (including singleton vectors), then cancels whatever packets he might have fully decoded (corresponding to singleton vectors), and moves on to repeating the exhaustive decoding attempts on the remaining set of packets (second iteration), etc. We observe that the third iteration provides negligible decoding progress if any.\footnote{For linear combination degrees \num{1} through \num{6}, the third iteration counts were \num{2}, \num{19}, \num{4}, \num{0}, \num{0}, \num{0}, respectively, and are thus barely visible on the histogram.} In turn, the mass of linear combinations counts (degrees \num{2} through \num{6} combined) is rather substantial when compared to degree-\num{1} counts, with degree-\num{2} combinations yielding highest counts. This hints at the significant potential of the PLNC approach.

\subsubsection{Frame-level decoding}

After the slot-level decoder has been run for all $T_\mathsf{f}$ slots that constitute a frame, a collection of decoded codeword combinations $\hat{\wv}_{\av^{(1)}}, \hat{\wv}_{\av^{(2)}}, \dotsc$ are available, with different weight vectors $\av^{(j)} = \{0,1\}^{|\mathcal{A}|}$. Let us stack these row vectors into a weight matrix
\begin{equation}
	\Am = \begin{bmatrix} \av^{(1)} \\ \av^{(2)} \\ \vdots \end{bmatrix}
\end{equation}
and denote $\mathcal{A}(i)$ the $i$-th ordered entry of the set of active users $\mathcal{A} \subseteq [L]$, e.g., $\mathcal{A} = \{2,4,5,\dotsc\}$ would yield $\mathcal{A}(1)=2$, $\mathcal{A}(2)=4$, $\mathcal{A}(3)=5$, etc.
Choosing the matrix of decoded linear combinations $\hat{\Wm} = [ \hat{\wv}_{\av^{(1)}}^\Tr , \hat{\wv}_{\av^{(2)}}^\Tr , \dotsc ]^\Tr$ as an estimator for the true linear combinations
\begin{equation}
	\Wm
	= \begin{bmatrix} \wv_{\av^{(1)}} \\ \wv_{\av^{(2)}} \\ \vdots \end{bmatrix}
	= \Am \begin{bmatrix} \uv_{\mathcal{A}(1)} \\ \uv_{\mathcal{A}(2)} \\ \vdots \end{bmatrix},
\end{equation}
the frame-level decoder attempts to recover the individual messages. If $\Am$ has rank $|\mathcal{A}|$, the linear system can be fully solved by means of
\begin{equation}
	\hat{\uv} = (\Am^\Tr\Am)^{-1}\Am^\Tr\hat{\Wm},
\end{equation}
all operations being performed in the binary field.

In general, even if the rank of $\Am$ is less than $|\mathcal{A}|$, the decoder may still be able to recover some {\em subset} of the codewords. For a systematic treatment of these situations, it is convenient to put $\Am$ into reduced row echelon form $\Am_{\mathrm{RREF}} = \Rm\Am$ by Gaussian elimination via elementary row operations described by a square full-rank matrix $\Rm$. In RREF, the first $1$ of each row (called the {\em pivot}) is located strictly to the right of the pivot of the previous row, and every pivot is the only non-zero entry in its corresponding column. For example,
\begin{equation}
	\Am_{\mathrm{RREF}}
	= \begin{bmatrix}
		1 & 1 & 0 & 0 & 1 & 0 & 0 \\
		0 & 0 & 1 & 0 & 0 & 0 & 0 \\
		0 & 0 & 0 & 1 & 1 & 0 & 1 \\
		0 & 0 & 0 & 0 & 0 & 1 & 0 \\
	\end{bmatrix}
\end{equation}
is in RREF. If in addition, a pivot is the only non-zero entry of its row, then the corresponding codeword can be recovered. In the above example, $\uv_{\mathcal{A}(3)}$ and $\uv_{\mathcal{A}(6)}$ can be recovered, whereas $\uv_{\mathcal{A}(1)}, \uv_{\mathcal{A}(2)}, \uv_{\mathcal{A}(4)}, \uv_{\mathcal{A}(5)}, \uv_{\mathcal{A}(7)}$ cannot.

\section{Decoding decisions via Deep Neural Networks}

The main difficulty in the SD step stems from the fact that the number of linear combinations to attempt increases exponentially with the degree of collision. As argued before, there are $2^{|\mathcal{T}'|}-|\mathcal{T}'|-1$ relevant combinations, which makes exhaustive trials impractical.

Unlike SIC, in which it is well known that users should be decoded in order of decreasing channel gains, for SD there is no such simple rule of thumb, nor any known analytical way to evaluate or sort the probabilities of decoding success. Thus, using a machine learning algorithm to \emph{learn} to predict which linear combinations can be reliably decoded appears as a promising approach to this problem. The machine learning approach is suitable in this context for several reasons: (i) the computation complexity is shifted to the training phase and allows for highly efficient real-time implementation, (ii) the structure of the problem will remain very similar when the number of users or the distribution of the channel changes, thus offering an opportunity to \emph{transfer} the trained models,\footnote{Note however that this aspect lies beyond the scope of this paper and will be investigated in future work} (iii) the function to be approximated is expected to be fairly smooth and well-behaved, and hence easy to learn.

We introduce a function $\pv = g(\hv)$ which takes the channel gains $h_\ell,\ \ell \in \mathcal{T}'$ as inputs and returns the vector $\pv$ containing the $2^{|\mathcal{T}'|}-1$ probabilities of successful decoding corresponding to each linear combination, i.e.,
\begin{equation}
	p_i = \mathsf{Pr}\bigl\{ \hat{\wv}_{\av(i)} = \wv_{\av(i)} \big| \hv \bigr\},
	\quad i = 1,\dotsc,2^{|\mathcal{T}'|}-1
\end{equation}
where $\av(i)$ denotes the length-$\mathcal{T}'$ binary expansion (as a row vector) of $i$.\footnote{Note that this representation of successful decoding probabilities only accounts for {\em marginal} probabilities, although the binary variates of the underlying random vector may {\em not} be independent. We limit this analysis to marginal probabilities for the sake of simplicity.}

If cognizant of $\pv$, the decoder will only attempt to decode the most promising combinations according to some heuristic. For instance, the decoder could choose those combinations $\av(i)$ whose probabilities $p_i$ lie above a given threshold $\tau$, or it could choose the topmost (in terms of high $p_i$) $\nu$ combinations (where $\nu$ is a fixed parameter) which are guaranteed to increase the rank of $\Am$ by $\nu$.

We will approximate $g$ using a DNN with parameter~$\bm{\theta}$ and we will denote the obtained function by~$g^{\bm{\theta}}$.
Beforehand, we will give a very concise introduction to supervised learning using DNNs in order to explain the general approach, before describing the training procedure in further detail.

\subsection{Refresher on supervised learning with DNNs}

A DNN is a function that can be expressed as the concatenation of several non-linear functions---so-called {\em layers}, where the $j$-th layer produces $n_j$ real-valued outputs (nodes). Each of these $n_j$ outputs is obtained from a so-called {\em neuron}, which takes a linear combination of the outputs of the previous layer, followed by the application of a non-linear \emph{activation function}. Specifically, let us denote the output of the $i$-th node of layer~$j$ by $y_i^j$ and the activation function by $\Phi$. The output $y_i^j$ is then given by
\begin{equation}
	y_i^j=\Phi\left( \sum_{i=1}^{n_{j-1}} \alpha_i^{j-1} y_i^{j-1} + \beta_i^{j-1} \right) 
\label{eq:intro_1}
\end{equation}
where $\alpha_i^j$ and $\beta_i^j$ are the so-called {\em weights} are {\em biases}, which constitute the parameters of the DNN (collectively referenced by $\bm{\theta}$) that need to be \emph{trained}.

In supervised learning, we aim to \emph{learn} a mapping $\xv_i \mapsto g(\xv_i)$ from a training data set~$\{(\xv_i,g(\xv_i))\}_{i=1,2,\dotsc}$. For supervised deep learning, we choose a neural network size (number of layers and nodes) and activation function(s) and then seek to adjust (train) parameters $\bm{\theta}$ so that $g^{\bm{\theta}}$ approximates $g$ well. Clearly, the input dimension of the first layer and the output dimension of the last layer have to be compatible with the input-output dimensions of the function $g$ to be approximated. The inner layers are referred to as {\em hidden layers}. Ideally, activation functions are chosen so as to reach the desired approximation accuracy at the fastest rate. A very popular activation function is the so-called rectified linear unit function
\begin{equation}
	\mathrm{ReLU}(z) = \max(z, 0)   \label{eq:intro_2}
\end{equation}
due to the simplicity of its derivative (which facilitates training via the backpropagation algorithm) and the fact that it has proven to be highly efficient in a large number of applications. In fact, DNNs have been known for many years but were notoriously difficult to train until recent breakthroughs both in terms of hardware and algorithmic efficiency of training~\cite{Lecun2015}.

It is important to understand that a DNN contains many parameters (so-called \emph{hyperparameters}) whose tuning is essential for an efficient training (e.g., number of nodes, layers, learning step, etc.). How to optimally design a DNN to learn a specific task is the focus of current research efforts in the field. In the present work, we show a new interesting application of DNNs and showcase a first working scheme. Yet, our design choices are led by a limited trial-and-error approach to improve the accuracy of training. Fine-tuning of the method herein presented will be the focus of future research.

\subsection{Training Parameters}

In the following, the function~$g^{\bm{\theta}}$ will be obtained from a DNN with \num{3} hidden layers each containing \num{50}~neurons each, initialized independently with samples from a zero-mean Gaussian distribution with variance $0.05$ to avoid saturation of the coefficients. For convenience, since we will fix the number of active users to \num{6} for our simulations, we train \num{6} separate DNNs, one for each degree of collision $d$ (from $1$ to $6$). Each DNN is trained with a training set of \num{e5}~channel realizations, each consisting of a real-valued vector~$\bm{h}_i \in \mathbb{R}^d,\ i = 1,\dotsc,10^5,\ d = 1,\dotsc,6$. The channel gains are distributed according to a Rician distribution with a Rician factor of $|\Exp[h]|^2/\Exp[|h|^2] = 0.9$. Each label in the training data consists of a binary vector~$\sv_i \in \{0,1\}^{2^d-1}$ which records success (one) or failure (zero) of one encoding--decoding operation for each of the $2^d-1$ possible linear codeword combinations. Hence our training data set is $\{(\hv_i,\sv_i)\}_{i=1}^{\num{e5}}$.

Note that this training set does not contain the output values of the function $g$ to be approximated. Instead, the training samples contain {\em realizations} of a random binary vector whose marginal probabilities are described by the outputs of $g$. By large sampling, we manage to approximate the function $g$.

We have implemented the DNN using the Tensorflow library which allows for an easy implementation of the DNN training, in particular thanks to automated differentiation routines. The training was carried out using the Adam-gradient descent algorithm with \num{e5} iterations and a gradient step size of $1.001$.

To assess the DNN's performance, we have computed the empirical frequencies of false alarm and missed detection (based on the training set statistics) as
\begin{subequations}
\begin{IEEEeqnarray}{rCl}
	\mathsf{P}_{\mathrm{FA}} &=& \frac{\bigl|\{ (i,j) \colon s_{ij} = 0 , g_j^{\bm{\theta}}(\hv_i) = 1 \}\bigr|}{\bigl|\{ (i,j) \colon s_{ij} = 0 \}\bigr|} \\
	\mathsf{P}_{\mathrm{MD}} &=& \frac{\bigl|\{ (i,j) \colon s_{ij} = 1 , g_j^{\bm{\theta}}(\hv_i) = 0 \}\bigr|}{\bigl|\{ (i,j) \colon s_{ij} = 1 \}\bigr|}
\end{IEEEeqnarray}
\end{subequations}
respectively. The frequencies obtained are shown in the following table.
\begin{table}[htp!]
\label{table_1}
\centering
\begin{tabular}{lll}
\hline
	& training & validation \\
\hline
	$\mathsf{P}_{\mathrm{FA}}$ & 6\% & 7 \% \\
\hline
	$\mathsf{P}_{\mathrm{MD}}$ & 18.21\% & 20.01\% \\
\hline
\end{tabular}
\end{table}

\begin{figure}[ht]
\begin{center}
\begin{tikzpicture}
\pgfplotstableread{coord/coord_frame_decoding.txt}\mydata
\pgfplotstableread{coord/coord_frame_decoding_DNN.txt}\mydataDNN
\pgfplotstableread{coord/coord_frame_decoding_SIC.txt}\mydataSIC
\begin{semilogyaxis}[
	legend style = {at = {(.05,.05)}, anchor=south west, legend columns=1},
	xlabel = {slots per frame},
	xtick = {0,1,...,20},
	xticklabels = {2,,,5,,,,,10,,,,,15,,,,,20},
	xmin = 0, xmax = 18,
	ymin = 0, ymax = 1,
	ylabel = {packet loss probability},
	height = 7cm, width = 8.5cm,
	]
\addplot[mark=o] table [x expr=\coordindex, y=2reps] {\mydata};
\addplot[mark=triangle] table [x expr=\coordindex, y=3reps, restrict x to domain=1:18] {\mydata};
\addplot[mark=square] table [x expr=\coordindex, y=4reps, restrict x to domain=2:18] {\mydata};
\addplot[dashed, mark=o] table [x expr=\coordindex, y=2reps] {\mydataDNN};
\addplot[dashed, mark=triangle] table [x expr=\coordindex, y=3reps, restrict x to domain=1:18] {\mydataDNN};
\addplot[dashed, mark=square] table [x expr=\coordindex, y=4reps, restrict x to domain=2:18] {\mydataDNN};
\addplot[densely dotted, mark=o] table [x expr=\coordindex, y=2reps] {\mydataSIC};
\addplot[densely dotted, mark=triangle] table [x expr=\coordindex, y=3reps, restrict x to domain=1:18] {\mydataSIC};
\addplot[densely dotted, mark=square] table [x expr=\coordindex, y=4reps, restrict x to domain=2:18] {\mydataSIC};
\legend{$r=2$, $r=3$, $r=4$}
\end{semilogyaxis}
\end{tikzpicture}
\end{center}
\caption{Packet loss probability of the PLNC coded random access scheme for varying number of repetitions $r$ and slots per frame, with exhaustive decoding attempts (solid curves), with DNN-aided decoding decisions (dashed curves) and with SIC decoding only (dotted curves).}
\label{fig:frame_decoding}
\end{figure}
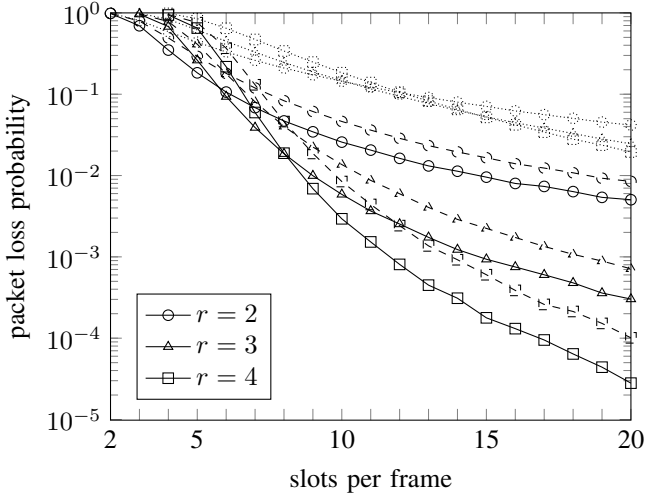

Note that false alarms lead to wasted decoding efforts, whereas missed detections negatively impact the system performance in terms of increased packet loss. In Figure~\ref{fig:frame_decoding}, we have plotted the packet loss probabilities resulting from exhaustive decoding against those obtained from a DNN-assisted receiver. The number of active users is fixed to $|\mathcal{A}|=6$ for all frames. The SNR is set to~\SI{10}{\decibel} and the fading distribution is Rician with factor 0.9, both for training data generation and in the validation phase.

As expected, the performance of a DNN-aided receiver does not fully match exhaustive search, but it still distinctly outperforms pure SIC decoding while keeping the number of decoding steps very low.

\section{Conclusion}

In this work, we have demonstrated how a machine learning algorithm based on deep learning can be a key enabler of a PLNC coded random access system by helping the receiver in making efficient decoding decisions. The main contribution of this work is to showcase the potential of this machine learning approach and to clarify the technical challenges still ahead in order to harvest the full benefits of this DNN-aided decoding approach. In extensions of this work, besides fine-tuning the deep learning setup, it will be interesting to account for more realistic conditions and factor in additional considerations, such as the robustness to parameter variations (SNR, fading distribution, imperfect channel knowledge), the adaptation of code rates and modulation schemes, retransmission requests, the scalability to very large collision degrees (in the context of massive machine-type communication), joint slot- and frame-level decoding decisions, etc. These are left for future investigation.

\bibliographystyle{IEEEtran}
\bibliography{IEEEabrv,references}

\end{document}